\documentclass[9pt,twocolumn,twoside]{opticajnl}
\journal{opticajournal} % use for journal or Optica Open submissions

\setboolean{shortarticle}{true}
\usepackage{graphicx}
\usepackage{lineno}
\title{Enhancing Faraday and Kerr rotations based on toroidal dipole mode in an all-dielectric magneto-optical metasurface}

\author[1]{Qin Tang}
\author[1]{Dandan Zhang}
\author[2,3,4]{Tingting Liu}
\author[1]{Wenxing Liu}
\author[1]{Qinghua Liao}
\author[1]{jizhou He}
\author[2,3]{Shuyuan Xiao}
\author[1,5]{Tianbao Yu}

\affil[1]{School of Physics and Materials Science, Nanchang University, Nanchang 330031, People’s Republic of China}
\affil[2]{Institute for Advanced Study, Nanchang University, Nanchang 330031, People’s Republic of China}
\affil[3]{Jiangxi Key Laboratory for Microscale Interdisciplinary Study, Nanchang University, Nanchang 330031, People’s Republic of China}

\affil[4]{liu.tt199104@163.com}
\affil[5]{yutianbao@ncu.edu.cn}
\begin{abstract}
The magneto-optical Faraday and Kerr effects are widely used in modern optical devices. In this letter, we propose an all-dielectric metasurface composed of perforated magneto-optical thin films, which can support the highly confined toroidal dipole resonance and provide full overlap between the localized electromagnetic field and the thin film, and consequently enhance the magneto-optical effects to an unprecedented degree. The numerical results based on finite element method show that the Faraday and Kerr rotations can reach -13.59° and 8.19° in the vicinity of toroidal dipole resonance, which are 21.2 and 32.8 times stronger than those in the equivalent thickness of thin films, respectively. In addition, we design an environment refractive index sensor based on the resonantly enhanced Faraday and Kerr rotations, with sensitivities of 62.96 nm/RIU and 73.16 nm/RIU, and the corresponding maximum figures of merit 132.22°/RIU and 429.45°/RIU, respectively. This work provides a new strategy for enhancing the magneto-optical effects at nanoscale, and paves the way for the research and development of magneto-optical metadevices such as sensors, memories, and circuits.
\end{abstract}

\setboolean{displaycopyright}{false} % Do not include copyright or licensing information in submission.

\begin{document}

\maketitle

Metasurfaces are essentially two-dimensional arrays of subwavelength optical resonators that are periodically arranged \cite{Zheludev2012,Xiao2020}. These structures allow for the precise control of electromagnetic wave characteristics such as amplitude, phase, and polarization \cite{Li2020,Liu2022}. While metasurfaces made of metal have been widely used in the past, those made of dielectric materials have recently gained popularity due to their high efficiency and diverse functionality \cite{Kuznetsov2016,Huang2023}. These dielectric metasurfaces can support various optical resonant modes, such as guided mode \cite{Wu2021}, magnetic dipole \cite{Tuz2020}, toroidal dipole \cite{Tuz2018}, anapole \cite{Yang2018}, and the recently discovered bound state in continuum \cite{Koshelev2018,Xiao2022,Qin2021}. The unique spectral features and significant field enhancement of dielectric metasurfaces can enhance the interaction between light and matter at nanoscale.

When light passes through a medium with a magnetic field, the polarization plane of the transmitted light will rotate, and the angle of rotation is called the Faraday rotation angle. In contrast, the Kerr rotation refers to the polarization rotation that occurs at the reflecting surface \cite{Qin2022}. Traditional methods often require using bulky magneto-optical material to achieve the desired Faraday and Kerr rotations, which make it difficult to achieve miniaturization and integration. However, for magneto-optical thin films, the Faraday and Kerr rotations are too weak. To sovle this dilemma, researchers have integrated plasmonic/dielectric nanostructures with magneto-optical thin films to enhance Faraday and Kerr rotations, particularly using resonance phenomena such as the Fano resonance \cite{Sadeghi2018,Guchhait2020}, localized surface plasmon resonance/surface plasmon resonance \cite{Chin2013,Floess2015,Wang2022,Nayak2021,Dmitriev2013,Lei2016,Belotelov2007}, and Mie resonance \cite{Xia2022}. However, these methods still face a problem of insufficient overlap between the localized electromagnetic field and the magneto-optic thin film. Recently, A. Christofi et al. \cite{Christofi2018} proposed a novel strategy to pattern magneto-optical thin film into nanodisks for fully utilizing the enhanced electromagnetic field within the structure and achieve a larger Faraday rotation. However, this discretely periodic arrangement of nanodisks embedded in the low-refractive-index medium background still presents significant challenges in experimental implementation.

To this end, we propose an all-dielectric metasurface composed of a uniform magneto-optical thin film, periodically perforated with hole arrays, deposited on a low-refractive-index substrate. This metasurface supports high-\emph{Q} toroidal dipole resonances, exhibiting extremely narrow spectral linewidths and strong field enhancement, which overlap well with the magneto-optical thin film and greatly enhance the magneto-optic effects. Numerical results show that in the proposed metasurface, the Faraday and Kerr rotations can reach -13.59° and 8.19° in the vicinity of toroidal dipole resonance, which are 21.2 and 32.76 times stronger than those in the equivalent thickness of thin films, respectively. As an exemplary application, we design an environment refractive index sensor based on the resonantly enhanced Faraday and Kerr rotations. The linear relationship between the wavelengths of zero Faraday rotation and Kerr rotation with refractive index is investigated, rendering sensitivities of 62.96 nm/RIU and 73.16 nm/RIU, and the corresponding maximum figures of merit (FoM) 132.22°/RIU and 429.45°/RIU, respectively. Our work exploits such simple perforated metasurface design, providing an integration-ready solution for magneto-optical devices and applications.

The proposed all-dielectric magneto-optical metasurface for enhancing Faraday and Kerr rotations is schematically shown in Figure 1. The unit cell is arranged in a square period of $p$ = 1000 nm, with a thickness of the magneto-optical thin film $t$ = 220 nm and a diameter of the hole $d$ = 200 nm. Since bismuth-substituted yttrium iron garnet (BIG) has high refractive index and low loss in the wavelength range of interest, it is assumed to be the magneto-optical material used in our study. If it is magnetized in the +\textit{z} direction, the dielectric tensor can be expressed as \cite{Christofi2018}:
\begin{equation}
\epsilon=\left(\begin{array}{ccc}
\varepsilon_r & i \varepsilon_i & 0 \\
-i \varepsilon_i & \varepsilon_r & 0 \\
0 & 0 & \varepsilon_r
\end{array}\right)
\label{eq:refname1}
\end{equation}
at saturation magnetization of about 0.05 T, the dielectric tensor components can be written as $\varepsilon_r = 6.25$ and $\varepsilon_i = 0.06$. The refractive index of silicon dioxide (Si$O_2$) is 1.45. The finite element method is used with commercially available software COMSOL Multiphysics to perform full-wave numerical simulations. These simulations involve an \textit{x}-polarized plane wave that is incident along the \textit{z} direction. Periodic boundary conditions are applied in both the \textit{x} and \textit{y} directions, while perfectly matched layers are used in the \textit{z} direction.
\begin{figure}[ht]
\centering
\includegraphics[width=8cm]{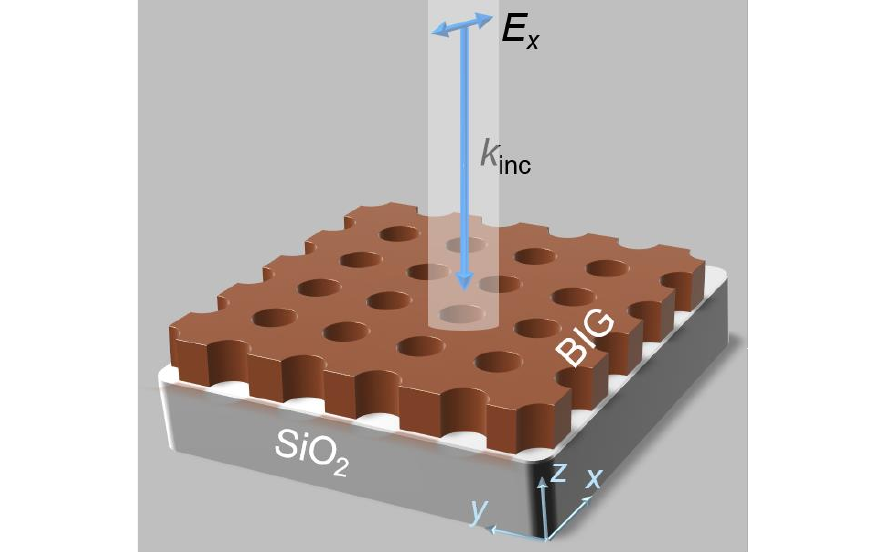}
\caption{The all-dielectric magneto-optical metasurface composed of a thin magnetic dielectric layer, periodically perforated with hole arrays, which is deposited on $\mathrm{SiO}_2$ substrate.}
\label{fig:false-color}
\end{figure}

Figure. 2(a) shows the simulated transmission spectrum (blue solid line), with an asymmetric line-shaped Fano resonance clearly observed near 1350.2 nm, which can be well-fitted by the classical Fano formula (red dashed line)
\begin{equation}
T(\omega)=\left|a_1+j a_2+\frac{b}{\omega-\omega_0+j \gamma}\right|^2,
\label{eq:refname1}
\end{equation}
here the fitting constants $a_1$, $a_2$, and $b$ are constant real numbers, $\omega_0$ is the resonant frequency, and $\gamma$ is the total leakage rate. The quality factor $Q$ is then calculated: $Q=\frac{\omega_0}{2 \gamma}$. The red dashed line in Fig. 2(a) describes the fitted curve with the detailed parameters: $a_1 = 0.33873$, $a_2 = 0.84036$, $b = 0.61187$, $\gamma = 1.66493\times 10^{12}$ Hz, $\omega_0 = 1.39626\times10^{15} Hz$, thus $Q = 419.31$. The simulation results are in good agreement with the fitting results. To reveal the physical origin of the Fano resonance, the near-field distribution at the resonant wavelength of 1350.2 nm is calculated, as shown in the inset of Fig. 2(a). A significant local electric field enhancement is observed in the magneto-optical metasurface, and the magnetic field, indicated by the black arrows, formed a closed loop in the \textit{y}-\textit{z} plane, indicating that the energy is strongly localized by the oscillation of the toroidal dipole moment in the \textit{x}-axis direction. Furthermore, we perform a multipole decomposition in the Cartesian coordinate system, where the induced multipole moments can be obtained by integrating the current density inside the metasurface unit cell \cite{Savinov2014}.
\begin{equation}
\vec{P}=\frac{1}{i \omega} \int \vec{j} d^3 r, \\
\label{eq:refname3}
\end{equation}
\begin{equation}
\vec{M}=\frac{1}{2 c} \int(\vec{r} \times\vec{j}) d^3 r, \\
\label{eq:refname3}
\end{equation}
\begin{equation}
\vec{T}=\frac{1}{10 c} \int\left[(\vec{r} \cdot \vec{j}) \vec{r}-2 r^2 \vec{j}\right] d^3 r, \\
\label{eq:refname3}
\end{equation}
\begin{equation}
Q_{\alpha \beta}^{(e)}=\frac{1}{2 i \omega} \int\left[r_\alpha j_\beta+r_\beta j_\alpha-\frac{2}{3}(\vec{r} \cdot \vec{j}) \delta_{\alpha, \beta}\right] d^3 r, \\
\label{eq:refname3}
\end{equation}
\begin{equation}
Q_{\alpha \beta}^{(m)}=\frac{1}{3 c} \int\left[(\vec{r} \times \vec{j})_\alpha r_\beta+\left[(\vec{r} \times \vec{j})_\beta r_\alpha\right]\right] d^3 r,
\label{eq:refname3}
\end{equation}
the symbols $c$ and $\omega$ epresent the speed of light and angular frequency of light, respectively. $r$ and $j$ represent the spatial position vector and current density, respectively. In the Cartesian coordinate system with $\alpha$, $\beta$ representing the $x$, $y$, and $z$ directions. $\vec{\mathrm{P}}$, $\vec{M}$, $\vec{T}$, $\vec{Q}^{(e)}$, and $\vec{Q}^{(m)}$ represent electric dipole (ED) moment, magnetic dipole (MD) moment, toroidal dipole (TD) moment, electric quadrupole (EQ) moment, and magnetic quadrupole (MQ) moment, respectively. The far-field scattered powers of these multipole moments are given by: $\textit{I}_p$=$\frac{2 \omega^4}{3 c^3}|\vec{P}|^2,$ $\textit{I}_M$=$\frac{2 \omega^4}{3 c^3}|\vec{M}|^2$, $\textit{I}_T$=$\frac{2 \omega^6}{3 c^5}|\vec{T}|^2$, $\textit{I}_{Q^{(e)}}$=$\frac{\omega^6}{5 \mathrm{c}^5} \sum \left|\vec{Q}_{\alpha, \beta}^{(\mathrm{e})}\right|^2$, $\textit{I}_{Q^{(m)}}$=$\frac{\omega^6}{40 \mathrm{c}^5} \sum\left|\vec{Q}_{\alpha, \beta}^{(m)}\right|^2$. Figure. 2(b) shows the multipole decomposition of the structure, where the far-field scattered power of the toroidal dipole is significantly enhanced at the resonance wavelength and dominates the scattered power in the far-field.

When a linearly polarized light beam propagates through a medium with a magnetic field, the polarization planes of the transmitted and reflected light are rotated relative to the original polarization plane, and the corresponding rotation angles are the Faraday rotation angle and the Kerr rotation angle, respectively. Then, the nonvanishing electric field components in the \textit{x} and \textit{y} directions appear.
The complex Faraday or Kerr rotation can be described by the following expressions:
\begin{equation}
\tilde{\theta}=\arctan \left(\frac{\tilde{\mathrm{E}}_{\mathrm{s}}}{\widetilde{\mathrm{E}}_{\mathrm{p}}}\right)=\theta+i \Phi,
\label{eq:refname3}
\end{equation}
here, $\theta$ and $\Phi$ represent rotation angle and ellipticity, respectively. Then, we
calculate the resonantly enhanced Faraday and Kerr rotations, and their ellipticity in the above magneto-optical metasurface, as shown in Figure 2 (c) and 2(d). $\theta_F$ and $\Phi_F$ represent the Faraday rotation angle and Faraday ellipticity, while $\theta_k$ and $\Phi_k$ represent the Kerr rotation angle and Kerr ellipticity, respectively. According to Figure 2(c), $\theta_F$ and $\Phi_F$ exhibit different trends as the wavelength changes. At wavelengths of 1350.1 nm, the Faraday rotation angle is -5.52°, which is unprecedented for films as thin as ~220 nm. Thanks to strongly confined local electromagnetic fields due to the toroidal dipole resonance, this value is increased by 8.3 times, compared to the unperforated BIG thin film with the equivalent thickness. Figure 2(d) shows the changes of $\theta_k$ and $\Phi_k$ with wavelength. At wavelengths of 1349.1 nm, the Kerr rotation angle is 3.58°, which is increased by 23.8 times, compared to the unperforated BIG film. It is worth noting that the ellipticity of the transmitted and reflected light increases in the vicinity of the resonance wavelength, but their positive and negative extremums appear at slightly different wavelengths. 
\begin{figure}[ht]
\centering
\includegraphics[width=9cm]{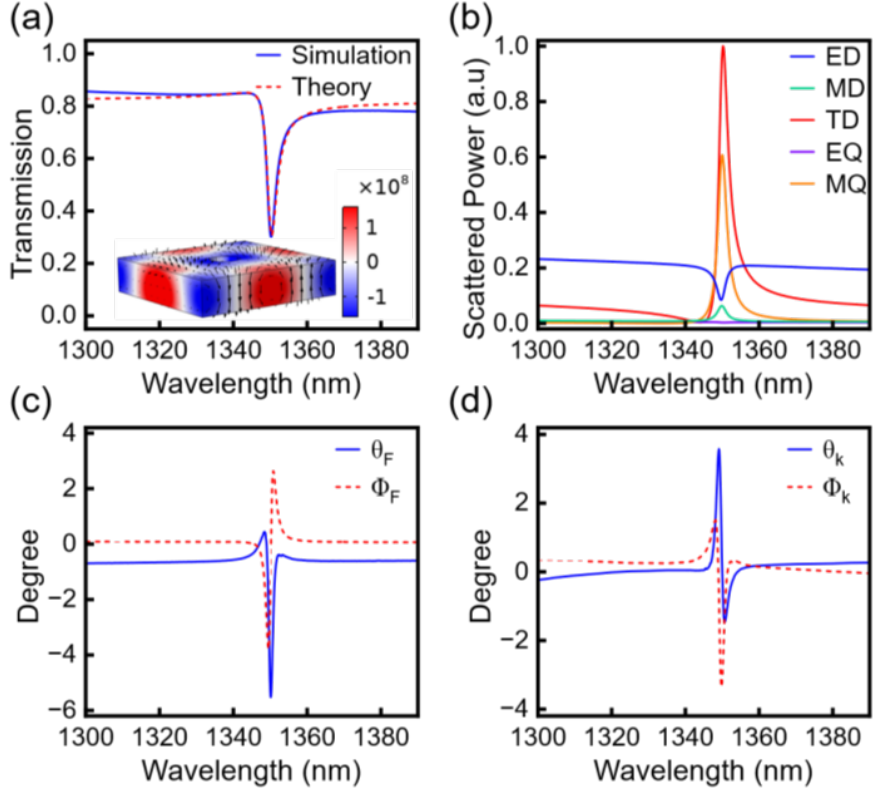}
\caption{(a) Transmission spectra of the proposed magneto-optical metasurface from numerical simulation and theoretical calculation, with the \textit{$E_x$} field distribution corresponding to the resonance wavelength shown in the inset, where the direction of the magnetic field distribution is indicated by the black arrow. (b) Multipole decomposition of the normalized scattering power in the Cartesian coordinate system, including electric dipole (ED), magnetic dipole (MD), toroidal dipole (TD), electric quadrupole (EQ), and magnetic quadrupole (MQ). (c) Faraday rotation (solid line) and ellipticity (dashed line), (d) Kerr rotation (solid line) and ellipticity (dashed line) of the magnetic-optical metasurface.}
\label{fig:false-color}
\end{figure}

We turn to the geometric dependence of the period, thickness, and air hole diameter of the magneto-optical metasurface on the Faraday and Kerr rotations. We first explore the variation in transmission, Faraday and Kerr rotations as a function of wavelength by adjusting the period. Figure 3(a) shows that as the period increases from 960 nm to 1020 nm, the toroidal dipole resonance shifts to longer wavelengths due to the increase of the effective refractive index, but the linewidth and quality factor at the resonance, which are highly related to the local field confinement, remain relatively stable. As a results, the Faraday and Kerr rotations show both redshifts to longer wavelengths, while the according maximum values of the rotation angles change slightly, as shown in Figure 3(b) and 3(c). The other two geometric parameters have much more significant effects. As can be seen in Figure 3(d), with the thickness increase from 200 nm to 240 nm, the resonance wavelength redshifts to longer wavelengths with a clearly observable narrowing resonance width, which reveals an increasingly confined local field to enhance the magneto-optical interaction. The concomitant Faraday and Kerr rotations exhibit the similar rising trendency, shown in Figure 3(e) and 3(f), in which the maximum values of Faraday and Kerr rotation angles remarkably reach -6.41° and 3.85°, respectively, in the thickest magneto-optical metasurface of 240 nm. On the contrary, the increase of the diameter of air hole would lead to the decrease of effective refractive index, and the resonance as well as the Faraday and Kerr rotations blueshift to shorter wavelengths. In Figure 3(g), 3(h), and 3(i), the maximum values of Faraday and Kerr rotation angles are obtained at the smallest diameter of 100 nm. -13.59° for Faraday rotation and 8.19° for Kerr rotation occur in the vicinity of toroidal dipole resonance, which are 21.2 and 32.8 times stronger than those in the equivalent thickness of unperforated BIG film, respectively.
\begin{figure}[ht]
\centering
\includegraphics[width=9cm]{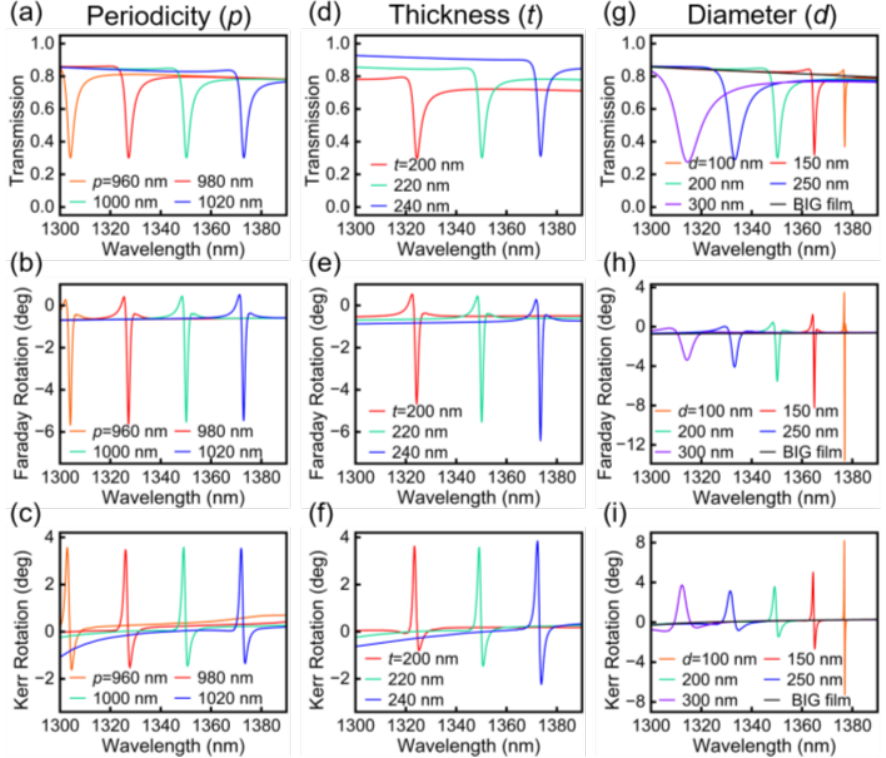}
\caption{Transmission spectra, Faraday, and Kerr rotations of magneto-optical metasurface with different periods, thicknesses, and diameters, The parameters for (a)-(c) are \textit{t} = 220 nm, \textit{d} = 200 nm, and different \textit{p}, the parameters for d-f are \textit{p} = 1000 nm, \textit{d} = 200 nm, and different \textit{t}, the parameters for g-i are \textit{p} = 1000 nm, \textit{t} = 220 nm, and different \textit{d}.}
\label{fig:false-color}
\end{figure}

Due to the significantly enhanced light-matter interaction by the toroidal dipole resonance, the magneto-optical Faraday and Kerr rotations in the proposed metasurface are further applied to the refractive index sensing to detect any tiny change in the environment. Without loss of generality, the geometric parameters are set as those in Figure 2. As shown in Figure 4(a) and 4(b), when the environmental refractive index $n_e$ gradually increases from 1.000 to 1.020, both Faraday and Kerr rotations shift towards longer wavelengths. Here we adopt the wavelength $\lambda_\theta$ of zero Faraday and Kerr rotations as the measuring point, as indicated by the black dashed lines. The sensitivity $S$ is thus defined as the ratio between the wavelength shift and refractive index change $S = d\lambda_\theta/dn_e$ \cite{Pourjamal2018, Zhu2020, Tang2018}. The FoM defined as FoM = $S\times|d\theta/d\lambda|$ is further introduced to evaluate the sensing performance, where $d\theta/d\lambda$ is the slope at wavelength $\lambda_\theta$. The results are summarized with linear fittings shown in Figure 4(c) and 4(d), in which $\lambda_\theta$ has a linear relationship with refractive index, with sensitivities of 62.96 nm/RIU and 73.16 nm/RIU, and the corresponding maximum figures of merit 132.22°/RIU and 429.45°/RIU, respectively for Faraday and Kerr rotations. Note that the sensitivity here is slightly lower than the plasmonic magnetic-optical metasurface \cite{Pourjamal2018}, which is mainly due to strongly localized field distribution within the all-dielectric metasurface structure. Therefore, the magneto-optical Faraday and Kerr rotations and the sensing performance have to be balanced when considering the structure design. 
\begin{figure}[ht]
\centering
\includegraphics[width=9cm]{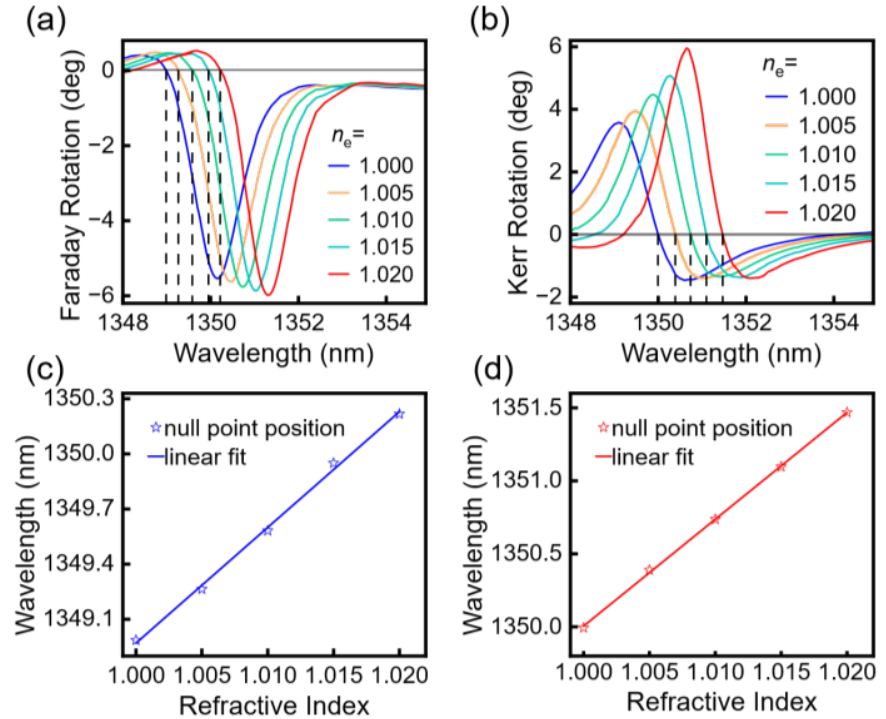}
\caption{Refractive index sensing using Faraday and Kerr rotations of magneto-optical metasurface. (a) Faraday rotation and (b) Kerr rotation as a function of the environmental refractive index. The black dashed vertical lines indicate the condition where the Faraday and Kerr rotations null at different environmental refractive indices. (c) Zero Faraday rotation wavelengths and (d) zero Kerr rotation wavelengths as a function of the environmental refractive index.}
\label{fig:false-color}
\end{figure}

In conclusion, we propose an all-dielectric metasurface composed of a perforated BIG film, aiming at enhancing the magneto-optical Faraday and Kerr rotations. The highly confined toroidal dipole resonance in the metasurface provides full overlap between the localized electromagnetic field and the thin film, which are confirmed by the electromagnetic field and current analysis as well as multipole decomposition method. The finite element simulation results show that the Faraday and Kerr rotations reach -13.59° and 8.19° in the vicinity of the resonance, which are 21.2 and 32.8 times stronger than those in the equivalent thickness of thin films, respectively. In addition, the structral dependence of the period, thickness, and air hole diameter of the magneto-optical metasurface on the magneto-optical effects are analyzed, revealing the tunability of the toroidal dipole resonance as well as the rotation angles in the wavelength of interest. Finally, the metasurface is applied to environmental refractive index sensing with sensitivities of 62.96 nm/RIU and 73.16 nm/RIU, and the corresponding FoM of 132.22°/RIU and 429.45°/RIU for Faraday and Kerr rotations, respectively. Our work provides a feasible solution to enhance Faraday and Kerr rotations, offering possibilities for smart designing of magneto-optical metadevices such as sensors, memories, and circuits.

\begin{backmatter}
\bmsection{Funding} National Natural Science Foundation of China (12064025, 11947065, 61901164, 12264028); Natural Science Foundation of Jiangxi Province (20212ACB202006, 20202BAB211007); Major Discipline Academic and Technical Leaders Training Program of Jiangxi Province (20204BCJ22012); Key Research and Development Program of Jiangxi Province (20192BBE50058).

\bmsection{Disclosures} The authors declare no conflicts of interest.

\bmsection{Data availability} Data underlying the results presented in this paper are
not publicly available at this time but may be obtained from the authors upon
reasonable request.

\end{backmatter}

% Bibliography
%\bibliography{refe}

\begin{thebibliography}{10}
	\newcommand{\enquote}[1]{``#1''}
	
	\bibitem{Zheludev2012}
	N.~I. Zheludev and Y.~S. Kivshar, {\protect\JournalTitle{Nature Materials}}
	\textbf{11}, 917 (2012).
	
	\bibitem{Xiao2020}
	S.~Xiao, T.~Wang, T.~Liu, C.~Zhou, X.~Jiang, and J.~Zhang,
	{\protect\JournalTitle{Journal of Physics D: Applied Physics}} \textbf{53},
	503002 (2020).
	
	\bibitem{Li2020}
	Z.~Li, C.~Chen, Z.~Guan, J.~Tao, S.~Chang, Q.~Dai, Y.~Xiao, Y.~Cui, Y.~Wang,
	S.~Yu, G.~Zheng, and S.~Zhang, {\protect\JournalTitle{Lasers {\&} Photonics
			Reviews}} \textbf{14}, 2000032 (2020).
	
	\bibitem{Liu2022}
	T.~Liu, Z.~Han, J.~Duan, and S.~Xiao, {\protect\JournalTitle{Physical Review
			Applied}} \textbf{18}, 044078 (2022).
	
	\bibitem{Kuznetsov2016}
	A.~I. Kuznetsov, A.~E. Miroshnichenko, M.~L. Brongersma, Y.~S. Kivshar, and
	B.~Luk'yanchuk, {\protect\JournalTitle{Science}} \textbf{354}, aag2472
	(2016).
	
	\bibitem{Huang2023}
	L.~Huang, L.~Xu, D.~A. Powell, W.~J. Padilla, and A.~E. Miroshnichenko,
	{\protect\JournalTitle{Physics Reports}} \textbf{1008}, 1 (2023).
	
	\bibitem{Wu2021}
	F.~Wu, M.~Luo, J.~Wu, C.~Fan, X.~Qi, Y.~Jian, D.~Liu, S.~Xiao, G.~Chen,
	H.~Jiang, Y.~Sun, and H.~Chen, {\protect\JournalTitle{Physical Review A}}
	\textbf{104}, 023518 (2021).
	
	\bibitem{Tuz2020}
	V.~R. Tuz, P.~Yu, V.~Dmitriev, and Y.~S. Kivshar,
	{\protect\JournalTitle{Physical Review Applied}} \textbf{13}, 044003 (2020).
	
	\bibitem{Tuz2018}
	V.~R. Tuz, V.~V. Khardikov, and Y.~S. Kivshar, {\protect\JournalTitle{{ACS}
			Photonics}} \textbf{5}, 1871 (2018).
	
	\bibitem{Yang2018}
	Y.~Yang, V.~A. Zenin, and S.~I. Bozhevolnyi, {\protect\JournalTitle{{ACS}
			Photonics}} \textbf{5}, 1960 (2018).
	
	\bibitem{Koshelev2018}
	K.~Koshelev, S.~Lepeshov, M.~Liu, A.~Bogdanov, and Y.~Kivshar,
	{\protect\JournalTitle{Physical Review Letters}} \textbf{121}, 193903 (2018).
	
	\bibitem{Xiao2022}
	S.~Xiao, M.~Qin, J.~Duan, F.~Wu, and T.~Liu, {\protect\JournalTitle{Physical
			Review B}} \textbf{105}, 195440 (2022).
	
	\bibitem{Qin2021}
	M.~Qin, S.~Xiao, W.~Liu, M.~Ouyang, T.~Yu, T.~Wang, and Q.~Liao,
	{\protect\JournalTitle{Optics Express}} \textbf{29}, 18026 (2021).
	
	\bibitem{Qin2022}
	J.~Qin, S.~Xia, W.~Yang, H.~Wang, W.~Yan, Y.~Yang, Z.~Wei, W.~Liu, Y.~Luo,
	L.~Deng, and L.~Bi, {\protect\JournalTitle{Nanophotonics}} \textbf{11}, 2639
	(2022).
	
	\bibitem{Sadeghi2018}
	S.~Sadeghi and S.~Hamidi, {\protect\JournalTitle{Journal of Magnetism and
			Magnetic Materials}} \textbf{451}, 305 (2018).
	
	\bibitem{Guchhait2020}
	S.~Guchhait, A.~B. S, N.~Modak, J.~K. Nayak, A.~Panda, M.~Pal, and N.~Ghosh,
	{\protect\JournalTitle{Scientific Reports}} \textbf{10}, 11464 (2020).
	
	\bibitem{Chin2013}
	J.~Y. Chin, T.~Steinle, T.~Wehlus, D.~Dregely, T.~Weiss, V.~I. Belotelov,
	B.~Stritzker, and H.~Giessen, {\protect\JournalTitle{Nature Communications}}
	\textbf{4}, 1599 (2013).
	
	\bibitem{Floess2015}
	D.~Floess, J.~Y. Chin, A.~Kawatani, D.~Dregely, H.-U. Habermeier, T.~Weiss, and
	H.~Giessen, {\protect\JournalTitle{Light: Science {\&} Applications}}
	\textbf{4}, e284 (2015).
	
	\bibitem{Wang2022}
	Z.~Wang, Z.~Wang, M.~Gao, L.~Kong, J.~Lan, J.~Zhao, P.~Long, J.~Kang, X.~Zheng,
	S.~Huang, and S.~Li, {\protect\JournalTitle{Optics Express}} \textbf{30},
	6700 (2022).
	
	\bibitem{Nayak2021}
	J.~K. Nayak, S.~Guchhait, A.~K. Singh, and N.~Ghosh,
	{\protect\JournalTitle{Communications Physics}} \textbf{4}, 102 (2021).
	
	\bibitem{Dmitriev2013}
	V.~Dmitriev, F.~Paix{\~{a}}o, and M.~Kawakatsu, {\protect\JournalTitle{Optics
			Letters}} \textbf{38}, 1052 (2013).
	
	\bibitem{Lei2016}
	C.~Lei, L.~Chen, Z.~Tang, D.~Li, Z.~Cheng, S.~Tang, and Y.~Du,
	{\protect\JournalTitle{Optics Letters}} \textbf{41}, 729 (2016).
	
	\bibitem{Belotelov2007}
	V.~I. Belotelov, L.~L. Doskolovich, and A.~K. Zvezdin,
	{\protect\JournalTitle{Physical Review Letters}} \textbf{98}, 077401 (2007).
	
	\bibitem{Xia2022}
	S.~Xia, D.~O. Ignatyeva, Q.~Liu, J.~Qin, T.~Kang, W.~Yang, Y.~Chen, H.~Duan,
	L.~Deng, D.~Long, M.~Veis, V.~I. Belotelov, and L.~Bi,
	{\protect\JournalTitle{{ACS} Photonics}} \textbf{9}, 1240 (2022).
	
	\bibitem{Christofi2018}
	A.~Christofi, Y.~Kawaguchi, A.~Al{\`{u}}, and A.~B. Khanikaev,
	{\protect\JournalTitle{Optics Letters}} \textbf{43}, 1838 (2018).
	
	\bibitem{Savinov2014}
	V.~Savinov, V.~A. Fedotov, and N.~I. Zheludev, {\protect\JournalTitle{Physical
			Review B}} \textbf{89}, 205112 (2014).
	
	\bibitem{Pourjamal2018}
	S.~Pourjamal, M.~Kataja, N.~Maccaferri, P.~Vavassori, and S.~van Dijken,
	{\protect\JournalTitle{Nanophotonics}} \textbf{7}, 905 (2018).
	
	\bibitem{Zhu2020}
	R.~Zhu, L.~Chen, S.~Wang, S.~Tang, and Y.~Du, {\protect\JournalTitle{Optics
			Letters}} \textbf{45}, 5872 (2020).
	
	\bibitem{Tang2018}
	Z.~Tang, L.~Chen, C.~Zhang, S.~Zhang, C.~Lei, D.~Li, S.~Wang, S.~Tang, and
	Y.~Du, {\protect\JournalTitle{Optics Letters}} \textbf{43}, 5090 (2018).
	
\end{thebibliography}

% Full bibliography added automatically for Optics Letters submissions; the following line will simply be ignored if submitting to other journals.
% Note that this extra page will not count against page length
%\bibliographyfullrefs{refe}

%Manual citation list
%\begin{thebibliography}{1}
%\bibitem{Zhang:14}
%Y.~Zhang, S.~Qiao, L.~Sun, Q.~W. Shi, W.~Huang, %L.~Li, and Z.~Yang,
 % \enquote{Photoinduced active terahertz metamaterials with nanostructured
  %vanadium dioxide film deposited by sol-gel method,} Opt. Express \textbf{22},
  %11070--11078 (2014).
%\end{thebibliography}

% Please include bios and photos of all authors for aop articles
\ifthenelse{\equal{\journalref}{aop}}{%
\section*{Author Biographies}
\begingroup
\setlength\intextsep{0pt}
\begin{minipage}[t][6.3cm][t]{1.0\textwidth} % Adjust height [6.3cm] as required for separation of bio photos.
  \begin{wrapfigure}{L}{0.25\textwidth}
    \includegraphics[width=0.25\textwidth]{john_smith.eps}
  \end{wrapfigure}
  \noindent
  {\bfseries John Smith} received his BSc (Mathematics) in 2000 from The University of Maryland. His research interests include lasers and optics.
\end{minipage}
\begin{minipage}{1.0\textwidth}
  \begin{wrapfigure}{L}{0.25\textwidth}
    \includegraphics[width=0.25\textwidth]{alice_smith.eps}
  \end{wrapfigure}
  \noindent
  {\bfseries Alice Smith} also received her BSc (Mathematics) in 2000 from The University of Maryland. Her research interests also include lasers and optics.
\end{minipage}
\endgroup
}{}

\end{document}